# Unveiling the Landscape of Smart Contract Vulnerabilities: A Detailed Examination and Codification of Vulnerabilities in Prominent Blockchains

Oualid Zaazaa[1] and Hanan El Bakkali[2]

Rabat IT Center, Smart Systems Laboratory (SSL), ENSIAS, Mohammed V University in Rabat, Rabat, Morocco

## ABSTRACT

*With the rise in using immature smart contract programming languages to build a decentralized application, more vulnerabilities have been introduced to the Blockchain and were the main reasons behind critical financial losses. Moreover, the immutability of Blockchain technology makes deployed smart contracts unfixable for the whole life of the Blockchain itself. The lack of complete and up-to-date resources that explain those vulnerabilities in detail has also contributed to increasing the number of vulnerabilities in Blockchain. In addition, the lack of a standardized nomination of the existing vulnerabilities has made redundant research and made developers more confused. Therefore, in this paper, we propose the most complete list of smart contract vulnerabilities that exist in the most popular Blockchains with a detailed explanation of each one of them. In addition, we propose a new codification system that facilitates the communication of those vulnerabilities between developers and researchers. This codification, help identify the most uncovered vulnerabilities to focus on in future research. Moreover, the discussed list of vulnerabilities covers multiple Blockchain and could be used for even future built Blockchains.*

## 1. INTRODUCTION

With the deployment of the Ethereum Blockchain for public use, new kinds of applications have been created. A Dapp or decentralized apps are the new generations of software that allow a full decentralization of both the data and the business logic. In most cases, those decentralized applications (Dapps) are built on what we call smart contracts. Smart contracts are simply pieces of software that run on each node with a final deterministic result.

This concept was built to remove the need for trusted third parties and allow freedom and security of transactions between peers. However, smart contracts are built by humans, and humans are known for making mistakes. Therefore, even if the whole idea of Blockchain is built on security, smart contracts are also vulnerable. Unfortunately, as a result of the immutability feature of the Blockchain environment, if a smart contract is deployed with a vulnerability, then the vulnerability will remain in the Blockchain forever. Moreover, most smart contract vulnerabilities have a dangerous impact on the business that usually leads to serious and direct financial losses. The most known attack was the Decentralized autonomous organization (DAO) [1], where a hacker exploited a known vulnerability called reentrancy to steal more than 150 million USD worth of ether (ETH) after a token sale priory conducted by the company.

However, the smart contract concept has opened the door for whole different use cases that are very useful to humanity. Moreover, more and more people have started to adopt this technology and built more complex applications to serve an objective. Unfortunately, not all the built smart contracts are secure by default, which is a normal result of being built by human and human make mistakes. Even worst, not all of them are audited to detect vulnerabilities before they get deployed on the main net and this is also clear in the Etherscan platform, as we can easily find a vulnerable smart contract deployed in the main net. Therefore, knowing and educating developers about those vulnerabilities is a must to help reduce the number of vulnerable smart contracts.

The problem with the smart contract security field is that it still not mature compared to classic applications. Therefore, there is no standard codification of all the possible vulnerabilities that could be found in a smart contract[2]. This lack of a standard codification allows for errors when naming vulnerabilities. For example, we were able to identify multiple situations where there two research have been performed on the same vulnerability but have been given a different names. This also creates redundancy in the works that could be performed by research around the world. Moreover, the same vulnerability could exist in multiple Blockchains but only given a code in one of them.

To solve this problem, in this paper we enumerate the most known vulnerabilities and Blockchains and proposed a more generic way of codifying the identified smart contracts that could be applied to any Blockchain.

The paper is organized as following, the first section discus the different key concepts required to understand the logic behind the vulnerabilities listed and explained in details in the followed section 4. The section 3 discus a new proposed way to share and communicate discovered vulnerabilities among the community of security research to avoid duplicate and misunderstandings. Section 5 ennumerate some of the eliminated issues discussed by the community due to multiple reasons. Finally, we finish our research by a conclusion around all the key aspects discussed in this paper.

## 2. BACKGROUND

The use of smart contracts in building Dapps is a rising field. Many projects are built and deployed in different Blockchains each day. Unfortunately, the technology used in building those apps is not mature enough compared to classic applications to hold and manage financial operations that once were only built with complex and heavy technologies that were tested multiple times before going to the production environment.

Therefore, the use of this technology by unaware developers and users, to build such apps has led to introducing critical vulnerabilities that lead directly to huge financial losses. This lack of knowledge came from the fact that this technology is still not well covered by research and there is a very few sources to learn about it. In this paper, we try to fill that gap by providing the most complete list of vulnerabilities that exists in the most popular Blockchains (Ethereum, VNT chain (Value Network), Hyperledger fabric, and EOSIO).

Other Blockchain eventually exists but most of them use the same technology as Ethereum to build a smart contract with small differences in other layers of the Blockchain (like consensus…).

The first Blockchain that offered the ability to build smart contracts was Ethereum and it was conceived in 2013 by Vitalik Buterin and went live in 2015. Ethereum is a permissionless type of Blockchain where anyone can become a node and it was based on proof of work until its big merge on 15 September 2022 when they fully switched to proof of stack consensus [3].

Hyperledger Fabric is an Enterprise grade permissioned distributed ledger platform that offers modularity and versatility for a broad set of industry use cases. Hyperledger was created under the supervision of the Linux Foundation, which has a long history of developing open-source projects with transparent governance and vibrant ecosystems. The key differentiation for this Blockchain is the ability to use different consensus protocols for each use case. In addition, Fabric can leverage consensus protocols that do not require a native cryptocurrency to incent costly mining or to fuel smart contract execution [4]. Smart contracts in hyperledger fabric Blockchain are called chaincode and are built using Java, Go, or Node.js.

As a blockchain platform, EOSIO is designed for enterprise-grade use cases and built for both public and private blockchain deployments. Building distributed applications on EOSIO follows familiar development patterns and programming languages used for developing non-blockchain applications (like C++)[5].

VNT chain could be seen as a hybrid Blockchain that combines the benefits of consortium and public Blockchains to offer companies more flexibility, security, and authority. This Blockchain is built on the vortex consensus and offers the ability to build a smart contract with either C/C++ or Rust [6]. To fill the gap of fewer resources to learn about vulnerabilities, multiple initiatives were made by the community. However, all of those initiatives either failed or suffer from limitations. The most popular and referenced initiative was the SWC-registry project [7] that lists multiple Ethereum smart contract vulnerabilities. The project covers multiple technical vulnerabilities and explains in detail each listed vulnerability with code examples for the vulnerability itself and the way to fix it. In addition, the project offers references to deeply analyze the vulnerability. Moreover, the project has linked each vulnerability to the class it belongs to according to the Common Weakness Enumeration (CWE) classification for better inerrability.

However, this project is still very limited as it only covers some vulnerabilities in the Ethereum Blockchain and none of the vulnerabilities that could be discovered in other Blockchain smart contracts like the ones discussed earlier in this paper. In addition, the project is outdated as the last update was made 2 years ago from the day this paper was written. Most vulnerabilities that could be discovered in the smart contract are way different than those in classic apps. Therefore, using only the CWE classification would make describing some vulnerabilities inaccurate.

## 3. CODIFICATION OF THE SMART CONTRACT VULNERABILITIES

After analyzing more than 100 published papers focusing on smart contract vulnerabilities, we have noticed many redundant kinds of research on vulnerabilities. Some research was discussing the same vulnerability but has given different names for it. As an example, in research made by Meng et al. a vulnerability known by timestamp dependencies was called "time manipulation" [8]. This redundancy confuses future researchers and makes it difficult for developers to understand the vulnerability itself.

To solve this issue we should unify the names given to vulnerabilities across all the research. To do that we propose a new codification specific for smart contract vulnerabilities, that cover all kind of Blockchain. The nomenclature proposed in this paper helps reduce redundancy and allows better communication among developers and researchers. In addition, the codification and the vulnerability naming proposed in this paper were built to be as generic as possible to be applicable for all available or future built Blockchains. The nomenclature was inspired by the CWE (Common weakness enumeration) and the SWC registry. The first part of the name "SWE" represents the Smart contract Weakness Enumeration. The second part is the ranking of the vulnerability and it is an incremental number. As the SWC registry nomenclature [7] is the most popular among the security research according to the papers selected in this research we decided to start the nomenclature from the same number as the SWC for better inerrability with the SWC registry. All the vulnerabilities' names were made generic, even for those that are only known for a specific Blockchain. The reason behind this choice is the fact that those vulnerabilities may appear in other Blockchains built in the future.

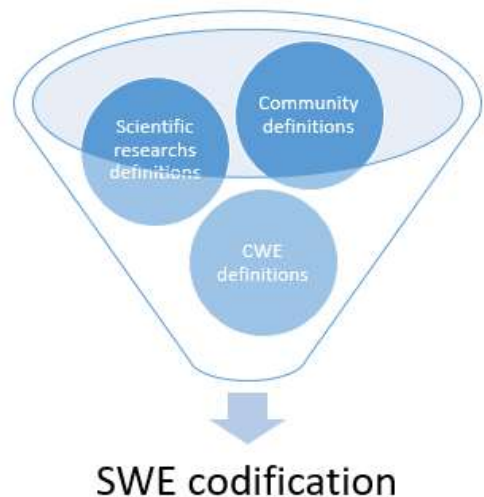

**Figure 1:** SWE codification combine all the previous works in a new unique and generic one.

## 4. SMART CONTRACT VULNERABILITIES

In this section, we are going to enumerate all the vulnerabilities that we were able to collect during our research, from other scientific papers and websites discussing specific scenarios of attacks. The code examples discussed in this section will use multiple smart contract programming languages depending on which Blockchain is more known by the vulnerability. However, the vulnerability concept remains the same even for the other vulnerabilities.

The impact discussed for each vulnerability is based on some real attacks that happened in the history of the Blockchain and on our analysis of the vulnerability itself. However, the impact may vary for each smart contract context and should be analyzed by the risk management team for better accuracy.

### 4.1 SWE-100: Function or state variable default visibility

Functions are public by default if a function visibility type is not provided. If a developer neglected to establish the visibility, then this might result in a vulnerability if the function should only be called internally for example [7], [9]–[15]. Vulnerable code [16] line (11) figure 2:

```
pragma solidity ^0.4.24;

contract HashForEther {

    function withdrawWinnings() {
        // Winner if the last 8 hex characters of the address are 0.
        require(uint32(msg.sender) == 0);
        _sendWinnings();
    }

    function _sendWinnings() {
        msg.sender.transfer(this.balance);
    }
```

**Figure 2:** Function or state variable default visibility vulnerable code

## 4.2 SWE-101: Integer Overflow and Underflow

When an arithmetic operation tries to produce a numeric value that is either higher than the maximum or lower than the lowest representable value, it is known as an integer overflow or integer underflow [7], [11]–[15], [17]–[20]. Vulnerable code line (7) figure 3:

```solidity
pragma solidity ^0.4.19;
contract IntegerOverflowMinimal {
    uint public count = 1;
    function run(uint256 input) public {
        count -= input;
    }
}
```

**Figure 3:** Integer Overflow and Underflow vulnerable

## 4.3 SWE-102: Outdated Compiler Version

Using an obsolete compiler version can introduce vulnerabilities that are already published and discussed by the community. [7], [15].

## 4.4 SWE-103: Floating compiler version

Using a fixed pragma helps to prevent the unintentional deployment of contracts using, for instance, an out-of-date compiler version that could bring flaws that have a detrimental impact on the contract system [7], [12], [13], [15]. Vulnerable code figure 4:

```solidity
pragma solidity >=0.4.0 < 0.6.0;
pragma solidity >=0.4.0<0.6.0;
pragma solidity >=0.4.14 <0.6.0;
pragma solidity >0.4.13 <0.6.0;
pragma solidity 0.4.24 - 0.5.2;
pragma solidity >=0.4.24 <=0.5.3 ~0.4.20;
pragma solidity <0.4.26;
pragma solidity ~0.4.20;
pragma solidity ^0.4.14;
pragma solidity 0.4.*;
pragma solidity 0.*;
pragma solidity *;
pragma solidity 0.4;
pragma solidity 0;
```

**Figure 4:** Floating compiler version examples

## 4.5 SWE-104: Unchecked Call Return Value

The succeeding program logic may behave unexpectedly if the call fails accidentally or if an attacker forces the call to fail. Therefore, the smart contract should check the result of an external call [7], [11]–[15], [17], [19], [20]. Vulnerable code figure 5:

```solidity
// SPDX-License-Identifier: GPL-3.0
pragma solidity 0.8.0;
contract vulContract{
  function createAccountInSmartContract2(address _to) payable public{
    _to.call{value: msg.value}("");
        // Continue the business logic
    }
}
```

**Figure 5:** Unchecked Call Return Value vulnerable

## 4.6 SWE-107: Reentrancy

Reentrancy attacks, also known as recursive call attacks, occur when a malicious contract calls into the calling contract before the function's first invocation has concluded. This might lead to undesired interactions between the many invocations of the function [7], [11]–[15], [17], [19]–[21].

Vulnerable code (12) figure 6:

```
pragma solidity 0.4.24;
contract Example {
  mapping (address => uint) public credit;
  function donate(address to) payable public{
    credit[to] += msg.value;
  }
  function withdraw(uint amount) public{
    if (credit[msg.sender]>= amount) {
      require(msg.sender.call.value(amount)());
      credit[msg.sender]-=amount;
    }
  }
}
```

```
  function queryCredit(address to) view public
returns(uint){
    return credit[to];
  }
}
```

**Figure 6:** Reentrancy vulnerable code

## 4.7 SWE-109: Uninitialized Storage Pointer

Variables in the local storage system that have not been initialized might point to unanticipated storage places in the contract, which may create vulnerabilities that are either are deliberate or accidental [7], [15], [20], [22].

Vulnerable code line (13) figure 7:

```
pragma solidity 0.4.0;
contract SecretContract {
    uint256 private secretNumber = 1;
    uint256 public oldSecretNumber;
    struct Game {
        address player;
        uint256 number;
    }
    function setGame() public{
        Game game;
        game.player = msg.sender;
        game.number = 1;
    }
}
```

**Figure 7:** Uninitialized Storage Pointer vulnerable

## 4.8 SWE-110: Assert Violation

Invariants are asserted using the Solidity assert() function [19]. A failed assert statement should never be encountered in properly working code. Therefore, if this happens then either the assert statement is improperly utilized to validate inputs as an example, or a issue exists in the smart contract that leads to an invalid state. The reason behind considering this as a vulnerability is that when it fails, the assert function consumes all the remaining gas and reverts the transaction. Contrarily, the required function reverts the transaction but transfer back the remaining gas. Therefore, the assert function should only be used to check invariants[7].

Vulnerable code line (5) figure 8:

```
pragma solidity ^0.4.19;

contract Example {
    function run(uint256 a, uint256 b) public {
        assert(a > b);
    uint256 t = a - b;
    }
}
```

**Figure 8:** Assert Violation vulnerable code

## 4.9 SWE-111: Use of Deprecated Solidity Functions

In Solidity, several operations and functions have been retired. Deprecated operators and functions may have security and compilation issues with newer versions of the Solidity compiler [7], [15], [20].

## 4.10 SWE-112: Delegatecall to Untrusted Callee

Delegating a call to a third-party contract is a common practice in smart contract building. This practice allows developers to avoid rewriting the same code by executing the one that is already on the chain but with the context of the caller. This practice is the main idea behind proxies. However, using the source code of other smart contracts is not always safe. If the callee contract is not checked and trusted, then this would include serious vulnerabilities in the smart contract. Moreover, if the callee smart contract is upgradable, then the business logic may change without realizing it, making the caller smart contract vulnerable [7], [11], [15], [20].

### 4.11 SWE-113: DoS with Failed Call

External calls may malfunction unintentionally or on purpose, which might result in a (Deny Of Service) DoS problem in the contract. It is preferable to execute each external call into a separate transaction so that the call recipient can start to reduce the harm brought on by such failures. This is especially true for payments, when it is preferable to let consumers withdraw cash rather than automatically send funds to them [7], [11]–[13], [15], [17], [20].Vulnerable code line (5) figure 9:

```solidity
pragma solidity ^0.8.0;
contract Example {
    // … code …
    function payAll() public {
        for(uint256 i;i<nbrUsers;i++)
        {
            payable(users[i]).transfer(0.1 ether);
        }
    }
}
```

**Figure 9:** DoS with Failed Call vulnerable code

### 4.12 SWE-114: Transaction Order Dependence

Blockchains that do not force a transaction order at which each one should be validated. Nodes usually choose the transaction with higher fees to maximize their gains. Therefore, if the smart contract depends on the order in which each transaction is validated, then a race condition could happen to create what we call in the Block-chain context a Transaction order dependence (TOD) vulnerability [7], [11], [13], [15], [17], [19], [20].
Vulnerable code line (5) figure 10:

```solidity
pragma solidity 0.8.0;

contract Example {
    uint256 secretAnswer = 100;
    function answer(uint256 a) public{
        if(a == secretAnswer)
        {
            payable(msg.sender).transfer(0.1 ether);
        }
    }
}
```

**Figure 10:** Transaction Order Dependence vulnerable

### 4.13 SWE-115: Authorization through tx.origin

One of the options that the solidity language offer to retrieve the address of the transaction sender  is by using the global variable tx.origin. However, if a malicious contract is called into by an authorized account, using the variable for permission might render a contract vulnerable. Since tx.origin returns the original sender of the transaction, in this instance the approved account, it is possible to bypass the authorization check [7], [11]–[13], [15], [17], [19], [20]. Vulnerable code line (8) figure 11:

```solidity
pragma solidity 0.4.24;
contract Example {
    address owner;
    function Example() public {
        owner = msg.sender;
    }
    function sendTo(address receiver, uint amount) public {
        require(tx.origin == owner);
        receiver.transfer(amount);
    }
}
```

**Figure 11:** Authorization through tx.origin vulnerable

### 4.14 SWE-116: Block values as a proxy for time

Access to time values is frequently required for contracts to carry out specific sorts of functionality. However, using the value of block.timestamp, or block.number is not a safe operation as those values could be manipulated by the nodes that execute the smart contract [7], [11]–[15], [17], [19]–[21], [23].

Vulnerable code line (5) figure 12:

```
pragma solidity 0.8.0;
contract Example {
  // ... code ...
  // Sale should finish exactly at 30 September 2022 18:00:00
  function isSaleFinished() public view returns (bool) {
    return block.timestamp >= 1664560800;
  }
    // ... code ...
}
```

**Figure 12:** Block values as a proxy for time

To fix this code, a developer should either use an Oracle system to check for the exact time or allow an error range. In the context of Ethereum Blockchain, a range of 900ms error could be applied as nodes are allowed to submit a value up to block.timestamp + 900. More than this value is considered wrong and leads to block not getting validated.

### 4.15 SWE-117: Signature Malleability
Although it is frequently assumed that each signature is unique when a cryptographic signature scheme is implemented in Ethereum contracts, signatures can be changed without having access to the private key and still be considered legitimate. One of the so-called "precompiled" contracts defined by the Ethereum virtual machine (EVM) standard is ecrecover, which implements elliptic curve public key recovery. The three values v, r, and s can be slightly altered by a malicious user to produce different valid signatures. If the signature is a component of the hash of the signed message, a system that verifies signatures at the contract level may be vulnerable to attacks. A malevolent user might produce valid signatures to replay previously signed communications [7].

### 4.16 SWE-118: Incorrect Constructor Name
During the deployment of a new smart contract, the only function that get called to initiate the different variables in the contract is the constructor function. Before Solidity version 0.4.22, the only method to define a constructor was to write a function that had the same name as the contract class that contained it. If a constructor function's name differs slightly from the contract name, it becomes a regular, callable function instead. This creates a vulnerability in the smart contract as anyone would call that function [7], [17], [20], [22]. Vulnerable code line (4) figure 13:

```
pragma solidity 0.4.0;
contract Example {
    address public owner;
  function example(){
      owner = msg.sender;
  }
  // ... code ...
}
```

**Figure 13:** Incorrect Constructor Name vulnerable code

### 4.17 SWE-119: Shadowing State Variables
Shadowing state variables happens when two contract, one inherit from the other and both have declared the same variable name. Or, a contract have a global variable with the same name as a local variable declared in a function. As a result, there would be two distinct copies of the same variable. This circumstance might go unreported in more intricate contract systems, which could result in security vulnerabilities. [7], [15], [22]. Vulnerable code line (9) Figure 14:

```
pragma solidity 0.4.0;
contract NFTSales {
    uint nftPrice = 0.1 ether;
    function getPrice() public constant returns(uint) {
        return nftPrice;
    }
}
contract PreNFTsale is NFTSales {
    uint nftPrice = 0.01 ether;
    function buy() public{
    uint256 price = getPrice();
  // ... code ...
    }
}
```

**Figure 14:** Shadowing State Variables vulnerable code

### 4.18 SWE-120: Weak Sources of Randomness from Chain Attributes

In a wide range of applications, having the ability to create random numbers is quite useful. However, it is difficult to build a reliable adequate source of randomness for apps in the Blockchain. In the Ethereum Blockchain for example smart contract using the blockhash block timestamp or any node-controlled global variable as a source of randomness is insecure as they can be controlled and known by the other smart contract in the same block [7], [11], [13]–[15], [17], [20], [21], [23]. Vulnerable code [24] lines (5,10,11,12) Figure 15:

```solidity
pragma solidity ^0.4.25;
contract RandomNumberGenerator {
    uint256 private salt =  block.timestamp;
    function random(uint max) view private returns (uint256 result) {
        // Get the best seed for randomness
        uint256 x = salt * 100 / max;
        uint256 y = salt * block.number / (salt % 5);
        uint256 seed = block.number/3+(salt % 300)+y;
        uint256 h = uint256(blockhash(seed));
        // Random number between 1 and max
        return uint256((h / x)) % max + 1;
    }
}
```

**Figure 15:** Weak Sources of Randomness from Chain Attributes vulnerable code

To get a good source of randomization and to keep the Blockchain result deterministic, developers should use Oracles.

### 4.19 SWE-121: Missing Protection against Signature Replay Attacks

To improve usability or reduce gas costs, signature verification may occasionally be required in smart contracts. By, maintaining a list of all digested message hashes and only allowing fresh message hashes to be processed, developers can prevent multiple processing of the same message hash supplied by different users. The SWE-121 and SWE-117 may seem to be the same but in the SWE-117 the attacker creates a valid signature with different information to bypass the verification. However, with SWE-121 the attacker only uses an existing and already executed transaction to make a replay attack [7], [11], [20].

### 4.20 SWE-122: Lack of Proper Signature Verification

Because of the flexibility and improved transferability that this offers, smart contract systems frequently permit users to sign messages off-chain rather than requiring them to do an on-chain transaction. Smart contract systems that handle signed messages must build their logic to determine the signed messages' authenticity before continuing to process them [15]. Smart contracts cannot directly engage with such systems since they are unable to sign messages, which is a constraint for them. Some signature verification systems make an effort to resolve this issue by presuming the authenticity of a signed message based on other techniques that do not have this restriction. An illustration of such a technique is to depend on the value of the msg.sender property and presume that if a signed message came from the sender address, then the sender address also produced it. In situations where proxies may be used to relay transactions, this may result in vulnerabilities[7], [25].

### 4.21 SWE-123: Requirement violation by the called smart contract

To validate external inputs the require() function is used by developers. However, inputs could be sent by smart contracts too. Therefore, a violation of such verification in that case is related to a bug in the contract that have provided the input or the condition specified in the require() function is too strong [7], [11], [15]. Vulnerable code line (14) Figure 16:

```solidity
pragma solidity 0.8.0;
contract Example {
    Example2 private f = new Example2();
    function getResult(int256 y) public view returns (int256) {
        return y * f.tenTimes(0);
    }
}
contract Example2 {
    function tenTimes(int256 x) public pure returns (int256) {
        require(x > 0);
        return 10 * x;
    }
}
```

**Figure 16:** Requirement violation by the called smart contract vulnerable code

## 4.22   SWE-124: Write to Arbitrary Storage Location

This vulnerability happens when a bug in a smart contract allows the attacker to override another one. In some cases, it could be the owner's address, which can help the attacker to bypass all the control management mechanisms [7], [15], [17], [20].

## 4.23   SWE-125: Incorrect Inheritance Order

Solidity, allow for the inheritance of multiple contracts by a single contract. Multiple inheritances pose an ambiguity known as the Diamond Problem: which base contract should be called in the child contract if two or more specifically the identical function? Therefore, the sequence of inheritance matters since base contracts have distinct priorities. This means that unexpected behaviour might result from ignoring inheritance order [7], [15], [20], [22].

## 4.24   SWE-126: Insufficient Gas Griefing

The insufficient gas griefing vulnerability happens when the smart contract does not check if it has enough gas to execute its sub-calls. The attacker could then exploit this vulnerability by sending just enough gas to execute the contract code but not its sub-calls. When the sub call does not get enough gas two scenarios could happen, either the smart contract revokes the whole transaction or it simply continues the execution. If the last scenario happens, then an unexpected behaviour could happen, leading to some serious financial losses [7], [11], [13], [15], [17], [21].
Vulnerable code line (5) Figure 17:

```
pragma solidity ^0.5.0;
contract Example {
    function createUser(TokenManager tokenManager, bytes memory _data) public returns(bool) {
        // ... code ...
    address(tokenManager).call(abi.encodeWithSignature("transferTokens(bytes)", _data));
        return true;
    }
}
// Contract called by Example
contract TokenManager {
    function transferTokens(bytes memory _data) public {
        // Execute contract code
    }
}
```

**Figure 17:** Insufficient Gas Griefing vulnerable code

## 4.25   SWE-127: Arbitrary Jump with Function Type Variable

Function types are supported by Solidity. In other words, a reference to a function with a matching signature can be given to a variable of a function type. It is possible to call the function stored to such a variable exactly like any other function. If the developer utilizes assembly instructions like "mstore" or the assign operator, an attacker might point a function type variable to any code instruction, breaking necessary validations and state changes [7], [20].
Vulnerable code [26] line (15) Figure 18:

```
pragma solidity ^0.4.25;

contract FunctionTypes {
    function withdraw() private {
        require(msg.value == 0, 'dont send funds!');
    address(msg.sender).transfer(address(this).balance);
    }
    function frwd() internal
        { withdraw(); }
    struct Func { function () internal f; }
    function breakIt() public payable {
        require(msg.value != 0, 'send funds!');
        Func memory func;
        func.f = frwd;
        assembly { mstore(func, add(mload(func), callvalue)) }
        func.f();
    }
}
```

**Figure 18:** Arbitrary Jump with Function Type Variable

## 4.26   SWE-128: DoS With Block Gas Limit

This vulnerability happens when the cost of running a function exceeds the block gas limit. Programming patterns that are safe in centralized apps might cause Denial of Service problems in smart contracts. Such a Denial of Service problem might result from altering an array of unknown sizes that grow in size over time [7], [11]–[13], [15], [17], [20].

Vulnerable code line (6) Figure 19:

```
pragma solidity ^0.4.25;

contract Example {
    uint256[] public usersBalances;
    function distributeRevenue() public returns (bool){
    for(uint i=0;i<usersBalances.length;i++) {
            usersBalances[i] += msg.value;
        }
        return true;
    }
}
```

**Figure 19:** DoS With Block Gas Limit vulnerable

### 4.27 SWE-129: Typographical Error

An example of a typographical error is when a declared operation unintentionally introduces a typo that also happens to be a legal operator (=+).This operator reinitiates the variable rather than adding a number to it. [7], [13], [15], [22].

Vulnerable code line (5) Figure 20:

```
pragma solidity ^0.4.25;
contract Example {
    uint8 n = 1;
    function exampleFunction() public {
        n =+ 1;
    }
}
```

**Figure 20:** Typographical Error vulnerable code

### 4.28 SWE-132: Unexpected balance

When contracts rigorously presume a particular Ether balance. The funds could be sent to smart contract with multiple techniques that cannot be all controlled by developers. For example the use selfdestruct, mine to the account, or forcefully deliver ether to a contract. In the worst-case situation, this can result in conditions called DOS that might make the contract useless [7], [12], [13], [20] or change push the smart contract to behave erroneously.

Vulnerable code line (7,8) Figure 21:

```
pragma solidity ^0.4.25;

contract Example {
    uint256 public totalFunds;
    mapping(address=>uint256) balances;
    function usersFunds() payable public returns (bool){
        balances[msg.sender] += msg.value;
        totalFunds += msg.value;
        return true;
    }
}
```

**Figure 21:** Unexpected balance vulnerable code

This smart contract tracks his balance and the balance of each user using a variable that only gets incremented using the usersFunds function. However, the attacker could send money to the contract without the need to pass through that function, which may have an impact on the business logic.

### 4.29 SWE-133: Hash collisions with multiple variable length arguments

Hash collision is possible when the abi.encodePacked() function is used with several parameters with different length. The abi.encodePacked() packs all items in order whether or not they are part of an array, elements could be transferred across arrays and it will still return the same encoding as long as all of the components are in the same order. An attacker might take advantage of this in a signature verification scenario by changing the arrangement of components in a prior function call to successfully evade permission [7].

Vulnerable code line (15) Figure 22:

```
pragma solidity ^0.5.0;

import "./ECDSA.sol";
contract AccessControl {
    using ECDSA for bytes32;
    mapping(address => bool) isAdmin;
    // ... code ...
    function addUsers(address[] calldata admins,
        address[] calldata regularUsers,
        bytes calldata signature) external
    {

        if (!isAdmin[msg.sender]) {
            bytes32 hash = keccak256(abi.encodePacked(admins, regularUsers));
            address signer = hash.toEthSignedMessageHash().recover(signature);
            require(isAdmin[signer], "Only admins can add users.");
        }
        // ... code ...
    }
}
```

**Figure 22:** Hash collisions with multiple variable length arguments vulnerable code

## 4.30   SWE-134: Message call with hardcoded gas amount

Hard forks may result in a considerable shift in the gas cost of EVM instructions, which might disrupt currently deployed contract systems that base their assumptions on constant gas prices. Therefore, hardcoding the amount of gas that a call may consume could in the future lead to DOS as the hard-coded gas amount might become insufficient [7], [27]. For example, Due to the SLOAD instruction's cost rise, the EIP 1884 damaged many existing smart contracts.

Vulnerable code line (10) Figure 23:

```
pragma solidity 0.8.0;

interface Users{
    function createUser() external;
}
contract Example1{
    function usersManagement(address c) public{
        Users u = Users(c);
        u.createUser{gas: 10000}();
        // ... code ...
    }
}
```

**Figure 23:** Message call with hardcoded gas amount

## 4.31   SWE-135: Code with no effects

Writing code in Solidity that doesn't have the desired results is possible. A "dead" code might be introduced, which is code that is unable to carry out its intended function. For instance, it's simple to forget the following parenthesis in the code msg.sender.call.value(address(this).balance)("");, which may result in a function running without sending money to msg.sender [7]. Vulnerable code line (9) Figure 24:

```
pragma solidity 0.6.0;
contract Example1{
    constructor() payable public{ }
    function withdraw() public onlyOwner{    msg.sender.call.value(address(this).balance);
    }
}
```

**Figure 24:** Code with no effects vulnerable code

## 4.32   SWE-136: Unencrypted Private Data On-Chain

The private type variable can be read. Attackers can examine transactions to identify values kept in the contract state [7], [11], [12], [15], [21]. Therefore, all the private data in the Blockchain have to be encrypted.

## 4.33   SWE-137: Access control management

Functions that are supposed to be executed by only users with specific roles are not protected. Therefore, any user can execute them. For example, functions that perform withdrawals or execute dangerous functions like

SELFDESTRUCT should be protected and only executed by users with the right permissions [11], [13]–[15], [19], [20].

### 4.34 SWE-138: Locked money

Smart contracts that implement payable functions but no withdrawal ones are vulnerable because they can lock money without the ability to get them back. Locking money in the smart contract could also be the result of a bug in the business logic of the smart contract [11], [12], [20].

### 4.35 SWE-139: Limited stack size

When the EVM's call stack reaches its maximum size of 1024 frames, the subsequent function call and all of its subcalls will fail. An attacker could take advantage of such a restriction by calling the vulnerable contract enough times to almost fill the stack, betting that the targeted contract would handle the resulting full stack failure incorrectly (or not at all), and then using the subsequent function call to take advantage of this weakness. We see that this call stack limit is essentially unreachable as a result of the modifications made as part of Ethereum's hard fork in October 2016 [15], [18], [21].

### 4.36 SWE-140: improper handling of exceptions

This vulnerability happens when the smart contract fails to handle possible exceptions that can be generated by the EVM like out of gas and others:

- Not having enough gas required for the execution
- Invalid instructions
- Insufficient stack items
- Invalid destination of jump opcodes
- Invalid stack size (greater than 1,024)

The out-of-gas vulnerability, for example, happens when the smart contract developer fix a certain amount of gas for an operation and does not expect it to get higher in the future. Therefore, the attacker may exploit this vulnerability to perform a malicious action [11], [13], [18], [20], [21], [28].

### 4.37 SWE-141: Dynamic library

This vulnerability happens when a smart contract depends on updatable components (other smart contracts). Therefore, if the attacker takes control over those contracts he can then change the contract logic [13].

### 4.38 SWE-142: Type cast

In Solidity, any address can be cast as a contract regardless of whether the code at the address represents the contract type being cast. This can be deceiving, especially when the author of the contract is trying to hide malicious code [11], [17], [21].

For example, if the programmer makes a mistake and calls another contract by mistake but it contains a function with the same structure expected by the compiler, the function will be executed and if the function does not exist, the function of fallback will be called. In any case, no exception will be launched [22].

### 4.39 SWE-143: Call to the unknown

This vulnerability happens when the smart contract calls another address without taking into consideration that this call may execute a fallback function in that address. This default behavior could cause some serious business logic vulnerabilities. Re-entrancy could be seen as a specific type of call to the unknown [12], [13], [15], [21].

Vulnerable code [29] line (16) Figure 25:

```
pragma solidity ^0.6.0;
contract King {
  //... code ...
  constructor() public payable {
    owner = msg.sender;
    king = msg.sender;
    prize = msg.value;
  }
  receive() external payable {
    require(msg.value >= prize || msg.sender==owner);
    king.transfer(msg.value);
    king = msg.sender;
    prize = msg.value;
  }
  function _king() public view returns (address payable) {
    return king;
  }
}
```

**Figure 25:** Call to the unknown vulnerable code

### 4.40 SWE-144: Assembly-based vulnerabilities

If the developer does not know exactly what he is doing, the direct usage of assembly language could lead to serious vulnerabilities [13], [15], [19].

### 4.41 SWE-145: Lost of ether

If the programmer enters an address to send ether and that address exists but it is an orphan address that belongs to no one then ether will be lost forever as there is no way to retrieve those ethers back [11], [21].

### 4.42 SWE-146: double constructor

The 0.4.22 version of the solidity compiler allows smart contracts to have a double constructor. This could lead to unexpected behavior in the contract [27]. Vulnerable code line (5,8) Figure 26:

```
pragma solidity 0.4.22;
contract Example{
    address public admin;
    function Example() public {
        admin = address(0xdeadbeef);
    }
    constructor() public {
        admin = msg.sender;
    }
}
```

**Figure 26:** Double constructor vulnerable code

### 4.43 SWE-147: Untrustworthy data feeds

The smart contract should use trustworthy oracles to gather information from the outside. For example, using centralized data feeds API calls, which may lead to making the contract unusable if they get shut down by the owner [15], [17].

### 4.44 SWE-148: improper or missing event handling

This vulnerability is usually related to the ERC-20 token where a transfer manually created function does not notify users of the success or failure of the transfer through an even. This is a critical flaw of the standard that led to the impossibility of error handling [9].

### 4.45 SWE-149: Forged Transfer Notification

The forged transfer Notification vulnerability happens when the smart contract fails to check if the destination (data.to) of the EOS transfer is itself within its transfer function. This may result in wrong consideration by the vulnerable smart contract as a successful EOS receive from the sender rather than just a notification [23]. Vulnerable code line (4) Figure 27:

```
class C: public eosio::contract {
  public: void transfer(uint64_t sender, uint64_t receiver) {
    auto data = unpack_action_data < st_transfer > ();
    if (data.from == _self) //no check for data.to
      return;
    doSomething();
  }
}
```

**Figure 27:** Forged Transfer Notification vulnearble

As shown in this example, if contract C fails to check whether the destination (i.e., data.to) of EOS transfer is itself within its transfer function, it may wrongly consider that it has received EOS from the sender rather than just a notification. As a result, it may wrongly credit the sender account, which has sent nothing to C.

### 4.46 SWE-150: Money leak

This vulnerability is generally related to a business logic bug that allows attackers to silently steal money from the contract [30]. Vulnerable code line (10,11) Figure 28:

```
pragma solidity 0.8.0;
contract Example{
  mapping(address => uint256) balances;
  // ... code ...
  function transferGains(address _to, uint256 _gains) onlyAdmin public{
      require(_to != address(0), "transfer to the zero address");
      uint256 fromBalance = balances[msg.sender];
      require(fromBalance >= _gains, "transfer amount exceeds balance");
      balances[msg.sender] = fromBalance - _gains*30/100;
      balances[_to] += _gains*30/100;
      emit Transfer(msg.sender, _to, _gains);
  }
  // ... code ...
}
```

**Figure 28:** Money leak vulnerable code

In this example, the admin sends 30% of the company gains to the specified user. The result of multiplying the gains per 30/100 is not always an integer. This means that in some situations the users will lose some money. This business logic bug is considered a vulnerability and is called a Money leak.

### 4.47 SWE-151: Unchecked division

This code is reserved for all the vulnerabilities that can arise from a non-checked division, like dividing by zero, or not taking into consideration the rest of the division [12], [13], [15]. Vulnerable code line (10,11) Figure 29:

```solidity
pragma solidity 0.8.0;
contract Example{
    mapping(address => uint256) balances;
    // ... code ...
    function transferGains(address _to, uint256 _gains) onlyAdmin public{
        require(_to != address(0), "transfer to the zero address");
        uint256 fromBalance = balances[msg.sender];
        require(fromBalance >= _gains, "transfer amount exceeds balance");
        balances[msg.sender] = fromBalance - _gains*30/100;
        balances[_to] += _gains*30/100;
        emit Transfer(msg.sender, _to, _gains);
    }
    // ... code ...
}
```

**Figure 29:** Unchecked division vulnerable code

### 4.48 SWE-152: Token API violation

Some smart contract-based tokens do not use standard smart contracts proposed by the community like openzeppelin ones. Therefore, this creates some serious vulnerabilities that may cause a critical impact on the token. For example, some tokens transfer function throws where they need to return a bool value [12]. Vulnerable code line (2) Figure 30:

```solidity
function transferFrom ( address _spender , uint _value ) returns ( bool success ) {
    require ( _value < 20 wei ) ;
    // ...
}
```

**Figure 30:** Token API violation vulnerable code.

### 4.49 SWE-153: Using components with known vulnerabilities

Using existing libraries and open-source codes in a newly created project is a very useful practice to avoid reinventing the wheel. However, open-source codes and libraries are not all safe and might be vulnerable to a public exploit. Therefore, even if the newly created smart contract is safe, those added codes may introduce hidden vulnerabilities [31].

### 4.50 SWE-154: Built-in symbol shadowing

This vulnerability happens when the developer makes use of a built-in symbol of the solidity language. This action could change the default behavior of the smart contract development language and have business logic bugs [20]. Vulnerable code line (1) Figure 31:

```solidity
modifier require(bool condition) {
    if (! condition) revert();
    _;
}
```

**Figure 31:** Built-in symbol shadowing vulnerable code.

### 4.51 SWE-155: Hardcoded addresses

Using a hardcoded address could make the smart contract useless in the future if one of those addresses did change [13]. Vulnerable code line (4,10) Figure 32:

```solidity
pragma solidity 0.8.0;
contract Example{
    address owner = 0x5B38Da6a701c568545dCfcB03FcB875f56beddC4;
    modifier onlyAdmin{
        require(msg.sender == owner);
        _;
    }
    function withdrawAll() onlyAdmin public{
payable(0x617F2E2fD72FD9D5503197092aC168c91465E7f2).transfer(address(this).balance);
    }
}
```

**Figure 32:** Hardcoded addresses vulnerable code.

In this example, if the address in line 4 or 10 gets lost (private key lost) these smart contract funds would be lost forever as there will be no way to withdraw them.

### 4.52 SWE-156: Send to zero address
This vulnerability happens when the smart contract does not check if the receiver is a Zero address [30].

### 4.53 SWE-157: Multiple calls in a single transaction
Performing calls inside a loop for example could create a DOS. If one call throws the whole execution stop. So if there is a loop on the user address and one of them is a contract that implements a fallback function that simply reverts [32]. Then a DOS attack will occur. Vulnerable code line (8) Figure 33:

```
pragma solidity 0.8.0;
contract Example{
    address[] public users;
    function withdrawAll() onlyAdmin public{
      for(uint256 i=0;i<users.length;i++)
      {
        payable(users[i]).transfer(1 ether);
      }
    }
}
```

**Figure 33:** Multiple calls in a single transaction vulnerable code.

### 4.54 SWE-158: Function clashing
To call a function in some Blockchains like Ethereum, the user send a function hash in the data field of his transaction. Then, based on that hash the EVM chose what function to execute. As the hash used in this operation is very weak, collisions are more likely to happen. The existence of the function clashing vulnerability could result in executing functions unconsciously, which then may lead to financial losses [20].

### 4.55 SWE-159: Business logic vulnerabilities
Business logic vulnerabilities are all the security bugs that result in financial losses and that are not related to the technology itself, but the design of the application and the different control mechanisms [20].

### 4.56 SWE-160: Identity verification
This vulnerability happens when the smart contract tries to verify if the caller is a wallet or a smart contract based on its code size. Unfortunately, this can be bypassed by calling the vulnerable contract in the constructor of the smart contract attacker, as the source code of the smart contract while executing the constructor is still 0 [14], [14].

### 4.57 SWE-161: Array length manipulation
For older versions of solidity, It is possible to manipulate the "length" field of an array with standard arithmetic functions [29]. A simple code to decrement the value of this field can bring the value to an underflow. "anArray.length--;" can make the array size set to max int. This vulnerability can disable a contract [13].

### 4.58 SWE-162: Non-determinism arising from Global Variable
Global variables are of great help for developers in all languages. However, those variables may change automatically [28], [33]. Therefore, using them in a Chaincode program may create non-determinism.

### 4.59 SWE-163: Non-determinism arising from KVS structure iteration
When developers use iteration with a map structure, the order of the key values is not unique due to the Go specification. As a result, nondeterminism may result from the employment of map structure iteration [28], [33].

### 4.60 SWE-164: Non-determinism arising from Reified Object Addresses
Through a pointer, programmers may control the value of the variables. A pointer is a memory address, and the address changes depending on the peer environment. As a result, using reified object addresses might lead to non-determinism [33].

### 4.61 SWE-165: Non-determinism arising from Concurrency of Program
Concurrency is well supported in Go utilizing goroutine and channel. Therefore, a race situation issue will quickly arise if a concurrent application is not handled properly [28], [33].

### 4.62 SWE-166: Non-determinism arising from Generating Random Number
Generating random numbers in the Go language is different than solidity. The Go language offers the right function to create an acceptable random number. However, using those functions result in multiple values across the peers, which then creates non-determinism [28], [33]. Vulnerable code line (4) Figure 34:

```
answer: = arg[0]
// Answer
rand.Seed(seed)
sel: = rand.Intn(10)

if answer == sel {
    SayGood(user, prize)
}
```

**Figure 34:** Non-determinism arising from Generating Random Number vulnerable code.

**4.63   SWE-167: Non-determinism arising from System Timestamp**

There is no assurance that timestamp routines will be called at the same moment in each endorsing peer, much like with random number generation, which creates non-determinism [28], [33]. Vulnerable code line (1) Figure 35:

```
sel: = now.Unix()
if time == sel {
    allowSelling(user, prize)
}
```

**Figure 35:** Non-determinism arising from System Timestamp vulnerable code.

**4.64   SWE-168: Non-determinism arising from Web service**

In classic applications, using a web service to retrieve a piece of information was the best solution. However, this act is not allowed in the context of Blockchain as each call to the web server could get a different result. In addition, using web services make the chaincode a centralized application which is against the Blockchain logic [28], [33].

**4.65   SWE-169: Non-determinism arising from System Command Execution**

In some use cases, executing a system command may be very helpful for a developer to avoid multiple lines of additional code. However, the result of the system command may change from node to node which creates non-determinism [28], [33].

**4.66   SWE-170: Non-determinism arising from External File Accessing**

Accessing system files does not usually give the same result from peer to peer. Therefore, performing such action in the chaincode may result in a non-determinism [28], [33].

**4.67   SWE-171: Non-determinism arising from External Library Calling**

Using libraries is a common practice in the software development process to avoid reinventing the wheel. However, some of those libraries may contain codes that create non-determinism [28], [33].

**4.68   SWE-172: Phantom read from range query**

In Hyperledger Fabric, accessing the state databases could be performed using the range methods such as GetQueryResult(), GetHistoryForKey(), and GetPrivateDataQueryResult(). Those methods need to be utilized carefully since they are not repeated throughout the validation step. Thus, phantom readings are not noticed. Phantom read alters the outcome of a procedure by reading data that previous transactions have added or removed [28], [33]. For example, after executing a chaincode that performs a range query and counts the number of elements, and writes the results to a key "count". If after the execution of the chaincode and just before the commit someone inserts a new element. Then without a range re-validation the transaction will write a wrong value of count.

**4.69   SWE-173: Field Declarations in chaincode structure**

When implementing chaincode in Go, programmers must implement two methods (Init() and Invoke()) to fulfill the chaincode interface. when the functions are implemented as methods of a structure. The structure's fields can be specified by developers. The methods provide access to the field [28], [33]. The field in the program can be utilized with a global state. However, because not every peer completes every transaction, the state is not maintained consistently among peers. Vulnerable code line (2) Figure 36:

```
type BadChaincode struct {
    globalValue string // this is a risk
}
func(t * BadChaincode) Invoke(stub shim.ChaincodeStubInterface) peer.Response {
    t.globalValue = args[0]
    return shim.Success([] byte("success"))
}
```

**Figure 36:** Field Declarations in chaincode structure vulnerable code

**4.70   SWE-174: Cross Channel Chaincode Invocation**

It is acceptable to execute a chaincode from another chaincode if they are both on the same channel. If not, then no data will be committed in the other channel and only the results of the chaincode function are accessible [33].

If both Chaincodes are on the same channel, it is okay to invoke a chaincode from another. If not, then no data will be committed in the other channel and developers only obtain what the chaincode function returns.

**4.71   SWE-175: Read-Write Conflict**

Read-Your-Write semantics are not supported by Hyperledger Fabric. Therefore, even if a transaction updates the value for a key before issuing a read, the value in the committed state is returned if a transaction reads a value for a key [28], [33]. Vulnerable code line (9) Figure 37:

```
// At the initial point: {key: "key", value: 0}
val: = 1
// Update the value from 0 to 1
err: = stub.PutState("key", val)
if err != nil {
  fmt.Printf("Error is happened. %s", err)
}
// The method returns 0, not 1
ret, err: = stub.GetState("key")
if err != nil {
  fmt.Printf("Error is happened. %s", err)
}
```

**Figure 37:** Read-Write Conflict vulnerable code.

### 4.72   SWE-176: Fake EOS Transfer

A smart contract on the EOSIO platform must use the eosio.token system smart contract to transfer EOS tokens to another smart contract. All smart contracts' EOS accounts are managed by the eosio.token system smart contract within its internal storage. The transfer function of the eosio.token system contract must be called by the sender contract during a transfer. The balances of the sender and receiver contract accounts will be changed appropriately within the eosio.token transfer function. While the transfer is being made, eosio.token will execute and require recipient() to alert both the sender and receiver contracts. A secure smart contract's apply function must make sure that eosio.token is the original receiver of the transfer operation (i.e., the code parameter of apply function). An attacker might conduct an inline call to the contract's transfer function to simulate an EOS transfer, though, if the contract is weak and does not verify that the code is eosio.token when the action is transfer. The susceptible contract may therefore mistakenly believe that the attacker has transferred EOS to it [23].
Vulnerable code line [5] Figure 38:

```
#define EOSIO_ABI_EX(TYPE, MEMBERS)
extern "C" {
  void apply(uint64_t receiver, uint64_t code, uint64_t action) {
    auto self = receiver;
    if (code == self || code == N(eosio.token) || action == N(onerror)) {
      TYPE thiscontract(self);
      switch (action) {
        EOSIO_API(TYPE, MEMBERS)
      }
    }
  }
}
```

**Figure 38:** Fake EOS Transfer vulnerable code.

As shown in the example above, within the apply function only checks whether the code is the contract itself or eosio.token (line 5) are performed, and no check whether the code is eosio.token when the action is transfer is done.

## 5.   ELIMINATED SMART CONTRACT VULNERABILITIES

Some of the discussed vulnerabilities in the selected papers were eliminated due to different reasons. The methodology followed to eliminate those issues was based on a personal idea of three smart contract security experts. Each one of the experts was giving separately a list of security issues to flag issues with no direct security impacts. Once all submitted their results we have performed a meeting with all of them and reached a consensus upon each one of their flagged issues.

The following list discusses the different eliminated issues and the reasons behind eliminating them:

### 5.1 Short address

This vulnerability is related to the apps that communicate with smart contracts. If the address is shorter than what is expected to be, then the behavior of the front-end application that communicates with the smart contract adds zeros which in some cases leads to business logic alteration [34].

### 5.2 Observability

The fact that everyone can read smart contracts is considered in the context of IoT as a vulnerability [17]. However, reading the source code of a smart contract is not a vulnerability it is a feature of Blockchain technology. Having access to the source code of a smart contract increase the trust between users and help reduce the number of vulnerabilities as they become analyzed by the community.

### 5.3 Immutability

Immutability is a feature of Blockchain technology. Some papers have considered that this feature represents a risk against the smart contract as a discovered vulnerability in a smart contract will remain there no matter what can be done. However, this has nothing to do with the smart contract code. In addition, this risk should be managed by the design team who chose to use the Blockchain as the technology to, support their project [17].

## 5.4 Unknown Unknowns

This is not a specific vulnerability in a smart contract but in any program in the world. Any smart contract technology could have vulnerabilities that are not yet discussed or published to the community. However, the risk of those kinds of vulnerabilities gets reduced while the technology gets more mature. Unfortunately, for newer technologies, this kind of vulnerability risk is still very high compared to Ethereum and the old app-building technologies [35].

## 5.5 Style guide violation

This is not a vulnerability as it does not impact, the execution of the smart contract. However, respecting the code style of the smart contract help auditors quickly understand the smart contract and quickly discover vulnerabilities [36].

## 5.6 Type inference

Type inference is the process of automatically detecting the variable types without explicit declaration in the source code. This automatic process is not always a vulnerability but could in some cases be the cause of other vulnerabilities like integer overflow and gas consumption [36].

## 5.7 Using multiple versions across different contracts

By mistake, developers often use different pragma versions across different files, which can cause inconsistency and introduce security issues. It is better to use one Solidity compiler version across all contracts instead of different versions with different bugs and security checks [32]. However, this is not a vulnerability if every contract is well written with specific security checks. Therefore, we cannot consider using multiple versions of smart contracts as a vulnerability.

## 5.8 Constant functions changing the state

This could not be considered a vulnerability as there is no impact on the contract where this issue exists [32]. All that could happen is related to the smart contract that will call this one.

## 5.9 Forwarding all remaining gas

Forwarding all the remaining gas to the called smart contract will only be a vulnerability if the smart contract is itself vulnerable to a reentrancy vulnerability [37]. Therefore, we cannot consider this as a separate vulnerability but it only makes the reentrancy vulnerability impact even worst.

## 5.10 Right-To-Left-Override control character (U+202E)

Right-to-left-override Unicode characters can be used maliciously to compel RTL text display and perplex users about a contract's true meaning. The idea here is to trick the user by displaying a source code that is different than what was displayed. For example, the attacker display in the Etherscan the source code with a Right-to-left-override control character to change the displayed source code logic and trick the user [38].

This technique could be used in building malicious smart contracts to steal users' money. However, the right-to-left-override control character does not create a technical vulnerability in the source code.

## 5.11 Presence of unused variables

In Solidity, unused variables are permitted and do not directly compromise security [32]. The recommended course of action is to avoid them since they can:

- increase calculations (and wasteful gas consumption)
- reveal defects or faulty data structures and are typically an indication of poor code quality
- produce code noise and reduce the readability of the code

# 6. CONCLUSIONS

The Dapp industry is still in the beginning and the technology behind it is still not mature compared to the classic development technologies. Therefore, finding vulnerabilities in their source code is a natural aspect. However, the impact of those vulnerabilities is usually disastrous to the business. Unfortunately, most of those vulnerabilities come from the lack of knowledge of the different vulnerabilities that developers could include in their source code. Moreover, most of the scientific research performed in this field has targeted the Ethereum Blockchain as it is the most popular and the first one to allow smart contract-based applications in its technology. For this reason and many others, we have decided to make scientific research to collect and classify all the vulnerabilities that could be discovered in smart contracts and that have been studied in any paper published from the day the smart contract concept was introduced to the world.

In addition, due to the inexistence of a standard nomination of the different vulnerabilities that could be found in a smart contract, a lot of research has mentioned the same vulnerability but with different naming. This problem has increased confusion among developers and even researchers. to address this issue in vulnerability data collection.

We have suggested creating a new system of naming and coding vulnerabilities to consolidate those with similar meanings that have been cited with different names.

Redundancy refers to the unnecessary repetition of information, which can lead to confusion and make it difficult for researchers to analyze and compare data. By consolidating vulnerabilities with similar meanings, the proposed system aims to reduce redundancy and increase clarity in vulnerability data collection. The new nomenclature and codification system is intended to be used as a standard for future vulnerability data collection, which would help to ensure consistency and comparability across different studies. This would make it easier for researchers to build on each other's work and avoid duplicating efforts. Overall, the proposed system would improve the efficiency and effectiveness of vulnerability research by reducing redundancy and confusion, and providing a standardized approach to naming and coding vulnerabilities.

Moreover, we have eliminated some of the issues mentioned in some researchs that have no direct security impact on the program itself or that cannot be solved or managed by the smart contract code. Our analysis has revealed that the current researches have primarily focused on only four blockchain technologies: Ethereum, Hyperledger Fabric, EOSIO, and VNT Chain. This represents a relatively limited set of technologies, with many other popular blockchain platforms remaining unexplored (such as Solana, Parity, and others). As a result, we intend to broaden the scope of our research in order to identify and elucidate smart contract vulnerabilities on these and other blockchain platforms, and develop novel techniques for their detection and resolution.

# 7. CONFLICTS OF INTEREST

The authors declare no conflict of interest.

# 8. REFERENCES

[1]    "The DAO: What Was the DAO Hack?," *Gemini*. https://www.gemini.com/cryptopedia/the-dao-hack-makerdao, https://www.gemini.com/cryptopedia/the-dao-hack-makerdao (accessed Sep. 23, 2022).

[2]    B. C. Gupta, N. Kumar, A. Handa, and S. K. Shukla, "An Insecurity Study of Ethereum Smart Contracts," in *Security, Privacy, and Applied Cryptography Engineering*, L. Batina, S. Picek, and M. Mondal, Eds., Cham: Springer International Publishing, 2020, pp. 188–207.

[3]    V. Buterin, "Ethereum: A Next-Generation Smart Contract and Decentralized Application Platform.," p. 36.

[4]    E. Androulaki *et al.*, "Hyperledger fabric: a distributed operating system for permissioned blockchains," in *Proceedings of the Thirteenth EuroSys Conference*, Porto Portugal: ACM, Apr. 2018, pp. 1–15. doi: 10.1145/3190508.3190538.

[5]    "Introduction To EOSIO." https://developers.eos.io/welcome/v2.2/introduction-to-eosio/index (accessed Sep. 22, 2022).

[6]    "VNT Chain." https://vntchain.io/ (accessed Sep. 22, 2022).

[7]    "Overview · Smart Contract Weakness Classification and Test Cases." http://swcregistry.io/ (accessed Sep. 30, 2022).

[8]    M. Ren *et al.*, "Empirical evaluation of smart contract testing: what is the best choice?," in *Proceedings of the 30th ACM SIGSOFT International Symposium on Software Testing and Analysis*, Virtual Denmark: ACM, Jul. 2021, pp. 566–579. doi: 10.1145/3460319.3464837.

[9]    Z. Gao, L. Jiang, X. Xia, D. Lo, and J. Grundy, "Checking Smart Contracts With Structural Code Embedding," *IEEE Trans. Software Eng.*, vol. 47, no. 12, pp. 2874–2891, Dec. 2021, doi: 10.1109/TSE.2020.2971482.

[10]   N. Ashizawa, N. Yanai, J. P. Cruz, and S. Okamura, "Eth2Vec: Learning Contract-Wide Code Representations for Vulnerability Detection on Ethereum Smart Contracts," in *Proceedings of the 3rd ACM International Symposium on Blockchain and Secure Critical Infrastructure*, Virtual Event Hong Kong: ACM, May 2021, pp. 47–59. doi: 10.1145/3457337.3457841.

[11]   M. Staderini, C. Palli, and A. Bondavalli, "Classification of Ethereum Vulnerabilities and their Propagations," in *2020 Second International Conference on Blockchain Computing and Applications (BCCA)*, Antalya, Turkey: IEEE, Nov. 2020, pp. 44–51. doi: 10.1109/BCCA50787.2020.9274458.

[12]   S. Tikhomirov, E. Voskresenskaya, I. Ivanitskiy, R. Takhaviev, E. Marchenko, and Y. Alexandrov, "SmartCheck: static analysis of ethereum smart contracts," in *Proceedings of the 1st International Workshop on Emerging Trends in Software Engineering for Blockchain*, Gothenburg Sweden: ACM, May 2018, pp. 9–16. doi: 10.1145/3194113.3194115.

[13] M. Demir, M. Alalfi, O. Turetken, and A. Ferworn, "Security Smells in Smart Contracts," in *2019 IEEE 19th International Conference on Software Quality, Reliability and Security Companion (QRS-C)*, Sofia, Bulgaria: IEEE, Jul. 2019, pp. 442–449. doi: 10.1109/QRS-C.2019.00086.

[14] D. He, Z. Deng, Y. Zhang, S. Chan, Y. Cheng, and N. Guizani, "Smart Contract Vulnerability Analysis and Security Audit," *IEEE Network*, vol. 34, no. 5, pp. 276–282, Sep. 2020, doi: 10.1109/MNET.001.1900656.

[15] L. Batina, S. Picek, and M. Mondal, Eds., *Security, Privacy, and Applied Cryptography Engineering: 10th International Conference, SPACE 2020, Kolkata, India, December 17–21, 2020, Proceedings*, vol. 12586. in Lecture Notes in Computer Science, vol. 12586. Cham: Springer International Publishing, 2020. doi: 10.1007/978-3-030-66626-2.

[16] *https://github.com/sigp/solidity-security-blog#visibility*. Sigma Prime, 2022. Accessed: Sep. 21, 2022. [Online]. Available: https://github.com/sigp/solidity-security-blog#visibility

[17] K. Peng, M. Li, H. Huang, C. Wang, S. Wan, and K.-K. R. Choo, "Security Challenges and Opportunities for Smart Contracts in Internet of Things: A Survey," *IEEE Internet Things J.*, vol. 8, no. 15, pp. 12004–12020, Aug. 2021, doi: 10.1109/JIOT.2021.3074544.

[18] I. Garfatta, K. Klai, W. Gaaloul, and M. Graiet, "A Survey on Formal Verification for Solidity Smart Contracts," in *2021 Australasian Computer Science Week Multiconference*, Dunedin New Zealand: ACM, Feb. 2021, pp. 1–10. doi: 10.1145/3437378.3437879.

[19] J.-W. Liao, T.-T. Tsai, C.-K. He, and C.-W. Tien, "SoliAudit: Smart Contract Vulnerability Assessment Based on Machine Learning and Fuzz Testing," in *2019 Sixth International Conference on Internet of Things: Systems, Management and Security (IOTSMS)*, Granada, Spain: IEEE, Oct. 2019, pp. 458–465. doi: 10.1109/IOTSMS48152.2019.8939256.

[20] B. Dia, N. Ivaki, and N. Laranjeiro, "An Empirical Evaluation of the Effectiveness of Smart Contract Verification Tools," in *2021 IEEE 26th Pacific Rim International Symposium on Dependable Computing (PRDC)*, Perth, Australia: IEEE, Dec. 2021, pp. 17–26. doi: 10.1109/PRDC53464.2021.00013.

[21] A. López Vivar, A. T. Castedo, A. L. Sandoval Orozco, and L. J. García Villalba, "An Analysis of Smart Contracts Security Threats Alongside Existing Solutions," *Entropy*, vol. 22, no. 2, p. 203, Feb. 2020, doi: 10.3390/e22020203.

[22] S. Hwang and S. Ryu, "Gap between theory and practice: an empirical study of security patches in solidity," in *Proceedings of the ACM/IEEE 42nd International Conference on Software Engineering*, Seoul South Korea: ACM, Jun. 2020, pp. 542–553. doi: 10.1145/3377811.3380424.

[23] D. Wang, B. Jiang, and W. K. Chan, "WANA: Symbolic Execution of Wasm Bytecode for Cross-Platform Smart Contract Vulnerability Detection*#*," p. 12.

[24] "theRun | Address 0xcac337492149bDB66b088bf5914beDfBf78cCC18 | Etherscan." https://etherscan.io/address/0xcac337492149bDB66b088bf5914beDfBf78cCC18#contracts (accessed Sep. 21, 2022).

[25] "Remove SignatureType.Caller by abandeali1 · Pull Request #1015 · 0xProject/0x-monorepo," *GitHub*. https://github.com/0xProject/0x-monorepo/pull/1015 (accessed Sep. 21, 2022).

[26] 262588213843476, "FunctionTypes.sol," *Gist*. https://gist.github.com/wadeAlexC/7a18de852693b3f890560ab6a211a2b8 (accessed Sep. 21, 2022).

[27] N. Matulevicius and L. C. Cordeiro, "Verifying Security Vulnerabilities for Blockchain-based Smart Contracts," in *2021 XI Brazilian Symposium on Computing Systems Engineering (SBESC)*, Florianopolis, Brazil: IEEE, Nov. 2021, pp. 1–8. doi: 10.1109/SBESC53686.2021.9628229.

[28] Y. Huang, Y. Bian, R. Li, J. L. Zhao, and P. Shi, "Smart Contract Security: A Software Lifecycle Perspective," *IEEE Access*, vol. 7, pp. 150184–150202, 2019, doi: 10.1109/ACCESS.2019.2946988.

[29] "Ethernaut." https://ethernaut.openzeppelin.com/level/0x43BA674B4fbb8B157b7441C2187bCdD2cdF84FD5 (accessed Sep. 21, 2022).

[30] T. A. Usman, A. A. Selcuk, and S. Ozarslan, "An Analysis of Ethereum Smart Contract Vulnerabilities," in *2021 International Conference on Information Security and Cryptology (ISCTURKEY)*, Ankara, Turkey: IEEE, Dec. 2021, pp. 99–104. doi: 10.1109/ISCTURKEY53027.2021.9654305.

[31] S. Ahmadjee, C. Mera-Gomez, and R. Bahsoon, "Assessing Smart Contracts Security Technical Debts," in *2021 IEEE/ACM International Conference on Technical Debt (TechDebt)*, Madrid, Spain: IEEE, May 2021, pp. 6–15. doi: 10.1109/TechDebt52882.2021.00010.

[32] P. Momeni, Y. Wang, and R. Samavi, "Machine Learning Model for Smart Contracts Security Analysis," in *2019 17th International Conference on Privacy, Security and Trust (PST)*, Fredericton, NB, Canada: IEEE, Aug. 2019, pp. 1–6. doi: 10.1109/PST47121.2019.8949045.

[33] K. Yamashita, Y. Nomura, E. Zhou, B. Pi, and S. Jun, "Potential Risks of Hyperledger Fabric Smart Contracts," in *2019 IEEE International Workshop on Blockchain Oriented Software Engineering (IWBOSE)*, Hangzhou, China: IEEE, Feb. 2019, pp. 1–10. doi: 10.1109/IWBOSE.2019.8666486.

[34] "Chapter 9: Smart Contract Security - Short Address/Parameter Attack - 《Mastering Ethereum》 - 书栈网 · BookStack." https://www.bookstack.cn/read/ethereumbook-en/spilt.10.c2a6b48ca6e1e33c.md (accessed Sep. 21, 2022).

[35] T. Durieux, J. F. Ferreira, R. Abreu, and P. Cruz, "Empirical review of automated analysis tools on 47,587 Ethereum smart contracts," in *Proceedings of the ACM/IEEE 42nd International Conference on Software Engineering*, Seoul South Korea: ACM, Jun. 2020, pp. 530–541. doi: 10.1145/3377811.3380364.

[36] P. Zhang, F. Xiao, and X. Luo, "SolidityCheck : Quickly Detecting Smart Contract Problems Through Regular Expressions." arXiv, Nov. 22, 2019. Accessed: Sep. 22, 2022. [Online]. Available: http://arxiv.org/abs/1911.09425

[37] S. Kim and S. Ryu, "Analysis of Blockchain Smart Contracts: Techniques and Insights," in *2020 IEEE Secure Development (SecDev)*, Atlanta, GA, USA: IEEE, Sep. 2020, pp. 65–73. doi: 10.1109/SecDev45635.2020.00026.

[38] A. Bouichou, S. Mezroui, and A. E. Oualkadi, "An overview of Ethereum and Solidity vulnerabilities," in *2020 International Symposium on Advanced Electrical and Communication Technologies (ISAECT)*, Marrakech, Morocco: IEEE, Nov. 2020, pp. 1–7. doi: 10.1109/ISAECT50560.2020.9523638.